\documentclass[11pt]{article}
\usepackage{moriond,epsfig}
\def\scs{\scriptscriptstyle}
\def\be{\begin{equation}}
\def\ee{\end{equation}}
\def\bea{\begin{eqnarray}}
\def\eea{\end{eqnarray}}

\newcommand{\f}{\frac}
\newcommand{\al}{\alpha_s}
\newcommand{\me}[1]{\langle#1\rangle}

\begin{document}
\vspace*{31mm}
\title{RADIATIVE $B$ DECAY IN THE SM$^\star$}

\author{M. MISIAK}

\address{Institute of Theoretical Physics, Warsaw University,\\
               Ho\.za 69, PL-00-681 Warsaw, Poland} 

\maketitle\abstracts{ 
The experimental and theoretical status of the inclusive decay
$\bar{B} \to X_s \gamma$ is briefly summarized. Results from a very
recent theoretical analysis are reported.  An ~$\sim\!\! 11\%$
increase in the SM prediction for BR$[\bar{B} \to X_s \gamma]$ is
found after replacing ~$m_c^{\rm pole}/m_b^{\rm pole}$~ by
~$m_c(\mu)/m_b^{1S}$~ in the NLO QCD correction. The well-known
enhancement of the branching ratio by QCD logarithms is identified as
an effect of $m_b$-evolution in the top-quark contribution to the
amplitude. This observation helps controlling the residual
scale-dependence. The present prediction for the ``total'' branching
ratio differs by 1.4$\sigma$ from the experimental world average.}

\noindent
The inclusive decay $\bar{B} \to X_s \gamma$ is well known as a good
testing ground for extensions of the SM. It arises mainly at one loop
in the SM, so it is naturally sensitive to electroweak-scale
exotica. All the parameters that are relevant for the SM prediction
are well measured in other processes. Moreover, there is no overall
non-perturbative factor in the theoretical expression for the decay
amplitude, contrary e.g. to the $B\bar{B}$ and $K\bar{K}$ mixing or to
$\bar{B}_s \to \mu^+ \mu^-$ that require lattice inputs at present. In
$\bar{B} \to X_s \gamma$ (within certain range of photon energy
cut-offs), non-perturbative effects enter only as corrections, in
analogy to the inclusive semileptonic decay $\bar{B} \to X_c e
\bar{\nu}_e$. Last but not least, the suppression of ${\rm
BR}_{\gamma} \equiv {\rm BR}[\bar{B} \to X_s \gamma]$ by \linebreak
$m_b/M_W \ll 1$ in the SM can be relaxed in many popular extensions of
the SM, e.g.  in the MSSM with large $\tan \beta$ or in the left-right
symmetric models. Then, the sensitivity of BR$_{\gamma}$ to exotic
particles extends much above the electroweak scale (up to $\Lambda
\sim M_W^2/m_b \simeq 1.3$~TeV), even if the CKM matrix remains the
only source of flavour violation.

Of course, the power of BR$_{\gamma}$ for testing new physics
crucially depends on how accurate its measurements are and how
accurate the theoretical prediction is. The current experimental
results read:
$( 3.03 \pm 0.47 ) \times 10^{-4}$ 
(CLEO$\,$\cite{CLEO1,CLEO2}), 
$(3.11 \pm 0.80_{\rm stat} \pm 0.72_{\rm sys} ) \times 10^{-4}$
(ALEPH$\,$\cite{ALEPH}),
$\left[ 3.36 \pm 0.53_{\rm stat} \pm 0.42_{\rm sys} 
~\left(+0.50\;-\!0.54\right)_{\rm model} \right] \times 10^{-4}$ 
(BELLE$\,$\cite{BELLE}).
Their weighted average 
\be \label{main.exp}
{\rm BR}_\gamma^{\rm exp} = (3.11 \pm 0.39 ) \times 10^{-4}
\ee
$\underline{\hspace{3cm}}$\\
$^\star${\footnotesize Contribution to the proceedings of the 
{\em XXXVIth Rencontres de Moriond}, Les Arcs, March 10--17, 2001.}

\begin{figure}
\begin{center}
\psfig{figure=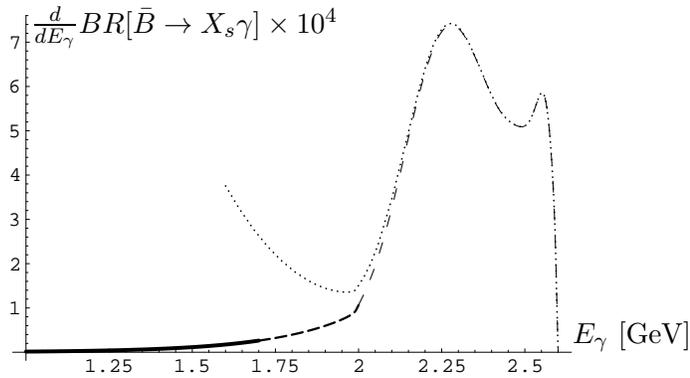,height=1.2in}\\[-34mm]
\hspace*{-29mm}
$\f{d}{dE_{\gamma}} BR[\bar{B} \to X_s \gamma] \times 10^4$\\[36mm]
\hspace*{9cm} $E_{\gamma}$~[GeV]\\[1mm]
\caption{An "artist view" of 
$\f{d}{dE_{\gamma}} BR[\bar{B} \to X_s \gamma]$.
\label{fig:spectrum}}
\end{center}
\vspace{-6mm}
\end{figure}

\newpage \noindent
has an error of around 13\%.  New results from CLEO, BELLE and BABAR
are expected soon. However, our limited knowledge of the photon
energy spectrum may restrict the accuracy of comparing theory with
experiment.  The $\bar{B} \to X_s \gamma$ photon spectrum in the
$\bar{B}$-meson rest frame is shown in Fig.~\ref{fig:spectrum}. The
solid and dashed lines describe the spectrum without the intermediate
$\psi$ contribution (i.e. the contribution from $\bar{B} \to X_s \psi$
followed by $\psi \to X' \gamma$). The dotted line shows how the
spectrum changes when the intermediate $\psi$ contribution is
included.\footnote{
It is a very rough estimate based on the measured 
 spectra of $\bar{B} \to X \psi$ and (boosted) $\psi \to X' \gamma$.}
This contribution has been effectively treated as background in all
the existing analyses of $\bar{B} \to X_s \gamma$, both on the
experimental and theoretical sides. This convention will be followed
below.

The thickness of the solid and dashed lines in Fig.~\ref{fig:spectrum}
reflects the degree of confidence with which the shape of the spectrum
is theoretically known. The prediction is quite solid where the line
is solid. For higher energies, it is only and ``artist view'' how the
spectrum could look like. We know that there is a peak there, and we
can determine the size of this peak, because the total inclusive decay
rate is calculable within the Heavy Quark Effective Theory. However,
the shape of the peak can be determined only experimentally. In this
respect, the recent results of CLEO$\,$\cite{CLEO1,CLEO2} are very
interesting. Unfortunately, their present energy cut-off \linebreak
$E_{\gamma} > 2$~GeV is still quite high.\footnote{
Moreover, it is imposed in the LAB frame, while 
$|E_{\gamma}^{\scs LAB} - E_{\gamma}^{\scs CM}|$ 
can reach 135~MeV for $E_{\gamma}^{\scs CM} \simeq 2$~GeV.}
Consequently, the present comparison of theory and experiment must
rely on a model-dependent extrapolation of the photon energy
spectrum.\cite{KN99} This issue might become less problematic once the
spectrum above the cut-off is more precisely measured.

In discussing the theoretical predictions below, I will assume that
the cut-off is already low enough, e.g.  $E_{\gamma} > 1.6$~GeV in the
$\bar{B}$-meson rest frame. In such a case, the dominant contribution
to BR$_\gamma$ is given by the partonic decay ~$b \to X_s \gamma$~ of
the $b$-quark. The electroweak one-loop diagrams that are relevant for
this decay were calculated 20 years ago. Seven years later, existence
of very large logarithmic QCD effects was realized. An enhancement of
~BR$_\gamma$~ by a factor of ~2.6~ (for $m_t = 175$~GeV) was found
after resummation of ~$\left( \al \ln M_W^2/m_b^2 \right)^n$~ to all
orders in $n$ with the help of renormalization-group techniques. Since
the perturbative uncertainties at LO were large, a calculation of NLO
QCD corrections was necessary. It was completed in 1996, up to small
two-loop matrix elements of the so-called penguin four-quark
operators. The NLO QCD corrections enhanced BR$_\gamma$ by another
$\sim$20\%. The electroweak and non-perturbative corrections that were
calculated later had smaller effects.

The overall uncertainty in the NLO prediction for BR$_\gamma$ is still
dominated by perturbative QCD. It has been estimated in several
papers.\cite{KN99,CMM97,BKP97,GM01} However, only the latter
article$\,$\cite{GM01} properly accounts for errors due to $m_c/m_b$.
In consequence, the predicted value of BR$_\gamma$ is significantly
higher than in the previous analyses. The uncertainty can be
maintained at the level of around 10\% thanks to an observation that
$m_b(\mu)$ in the top-quark contribution to the decay amplitude is the
main source of logarithmic QCD effects. Below, I will discuss those
very recent developments.

\begin{figure}
\begin{center}
\psfig{figure=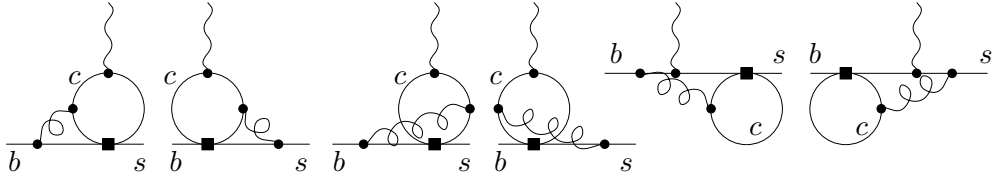,height=0.8in}
\caption{ Leading contributions to the matrix element
$\me{s \gamma | (\bar{s}c)_{\scs V-A}(\bar{c}b)_{\scs V-A} | b}$.}
\label{fig:ghw.diag}
\end{center}
\ \\[-31mm]
\hspace*{95mm}$b$\hspace{ 2cm}$s$\hspace{ 3mm}$b$\hspace{21mm}$s$\\[-2mm]
\hspace*{23mm}$c$\hspace{11mm}$c$\hspace{29mm}$c$\hspace{11mm}$c$\\[2mm]
\hspace*{113.5mm}$c$\hspace{12.5mm}$c$\\
\hspace*{15mm}$b$\hspace{15mm}$s$\hspace{3mm}$b$\hspace{15mm}$s$
 \hspace{2mm}$b$\hspace{15mm}$s$\hspace{3mm}$b$\hspace{15mm}$s$\\[-2mm]
\end{figure}

	The $(m_c/m_b)$-dependence of the $b \to s \gamma$ amplitude
arises in the diagrams shown in Fig.~\ref{fig:ghw.diag}, where the
$W$-boson propagator has been contracted to a point. We have to ask
what renormalization scheme should be used for quark masses. Should we
use
~$m_c^{\rm pole}/m_b^{\rm pole} = 0.29 \pm 0.02$~ 
or 
~$m_c^{\overline{\rm MS}}(\mu)/m_b^{\rm pole} \approx 0.22 \pm 0.04$~ 
(with $\mu \in [m_c,m_b]$)? In principle, such a question is a NNLO
issue, i.e. it is as relevant as three-loop corrections to the
considered diagrams. However, it is numerically very important,
because changing $m_c/m_b$ from 0.29 to 0.22 implies an increase of
BR$_\gamma$ by 11\%, i.e. by as much as the present experimental and
theoretical uncertainties.

	Since calculating three-loop corrections to
Fig.~\ref{fig:ghw.diag} would be a very difficult task at present, we
have to guess what the optimal choice of $m_c$ and $m_b$ is, on the
basis of our experience from other calculations. All the factors of
$m_c$ in Fig.~\ref{fig:ghw.diag} originate from explicit mass factors
in the charm-quark propagators. In the real part of the considered
amplitude, those charm quarks are dominantly off-shell, with momentum
scale $\mu$ set by some sizeable fraction of $m_b$. Therefore, it
seems reasonable to vary $\mu$ between $m_c \sim \f{1}{3}m_b$ and
$m_b$, and to use $m_c^{\overline{\rm MS}}(\mu)$ in the ratio
$m_c/m_b$.

        Factors of $m_b$ in Fig.~\ref{fig:ghw.diag} originate either
from the overall momentum release in $b \to s \gamma$ or from the
explicit appearance of $m_b$ in the $b$-quark propagators. In the
first case, the appropriate choice of $m_b$ is a low-virtuality mass.
In the second case, there is no intuitive argument that could tell us
whether $m_b^{\rm pole}$ or $m_b(m_b)$ is preferred. However, so long
as the three-loop diagrams remain unknown, setting all the factors of
$m_b$ equal to $m_b^{\rm pole}$ seems to be a good choice. Even a
better choice is the so-called 1S-mass of the b-quark defined as half
of the perturbative contribution to the $\Upsilon$ mass.  It is
leading-renormalon free and differs from $m_b^{\rm pole}$ only by 1\%
at one loop.

Once $m_c^{\overline{\rm MS}}(\mu)/m_b^{1S}$ with $\mu \in
[m_c,m_b]$ is used in Fig.~\ref{fig:ghw.diag}, the uncertainty in
BR$_\gamma$ significantly increases.  This is due in part to a strong
scale-dependence of $m_c(\mu)$. Moreover, in all the previous
analyses, the $m_c$-dependence of $\Gamma[b \to s \gamma]$ cancelled
partially against that of the semileptonic decay rate that is
conventionally used for normalization.  Once the different nature of
the charm mass in the two cases is appreciated, the cancellation no
longer takes place.

	Fortunately, it is possible to make several improvements in
the calculation, which allows us to maintain the theoretical
uncertainty at the level of around 10\%. In particular, good control
over the behaviour of QCD perturbation series is achieved by splitting
the charm- and top-quark-loop contributions to the $b \to s \gamma$
amplitude.  The overall factor of $m_b$ is frozen at the electroweak
scale in the top contribution to the effective vertex
~$m_b (\bar{s}_L \sigma^{\mu \nu} b_R) F_{\mu \nu}$.~
All the remaining factors of $m_b$ are expressed in terms of the
bottom 1S-mass.  

Splitting the charm and top contributions to the amplitude allows us
to better understand the origin of the well-known factor of
$\;\sim\!\! 3$ enhancement of BR$_\gamma$ by QCD logarithms. The charm
contribution is found to be extremely stable under logarithmic QCD
effects.  The QCD enhancement of the branching ratio appears to be
almost entirely due to the $\mu$-dependence of $m_b(\mu)$ in the
top-quark sector.

BR$_\gamma$ with an energy cut-off $E_0$ in the $\bar{B}$-meson rest
frame can be expressed as follows:
\be \label{main}
{\rm BR}[\bar{B} \to X_s \gamma]^{{\rm subtracted~} \psi,\;\psi'
}_{E_{\gamma} > E_0}
= {\rm BR}[\bar{B} \to X_c e \bar{\nu}]_{\rm exp} 
\left| \f{ V^*_{ts} V_{tb}}{V_{cb}} \right|^2 
\f{6 \alpha_{\rm em}}{\pi\;C} 
\left[ P(E_0) + N(E_0) \right],
\ee
where  
~$C = \left| V_{ub}/V_{cb} \right|^2  \Gamma[\bar{B} \to X_c e \bar{\nu}] / 
\Gamma[\bar{B} \to X_u e \bar{\nu}] \approx 0.575$~
is the phase-space factor for $\bar{B} \to X_c e \bar{\nu}$.
$N(E_0)$ is the non-perturbative correction. The perturbative
quantity $P(E_0)$ reads
\be \label{pert.ratio}
P(E_0) ~=~ 
\left| \f{V_{ub}}{ V^*_{ts} V_{tb}} \right|^2 
\f{\pi}{6 \alpha_{\rm em}}
\f{\Gamma[ b \to X_s \gamma]_{E_{\gamma} > E_0}}{
     \Gamma[ b \to X_u e \bar{\nu}]} ~=~ 
\left| K_c + r(m_t) K_t + \varepsilon_{\rm ew} \right|^2 + B(E_0).
\ee
Here, $K_t$ contains the top contributions to the ~$b \to s \gamma$~
amplitude. $K_c$ contains the remaining contributions, among which the
charm loops are by far dominant. The electroweak correction is denoted
by $\varepsilon_{\rm ew}$. The ratio
~$r(m_t) = m_b^{\overline{\rm MS}}(m_t)/m_b^{1S} \approx 0.578$~
appears in Eq.~(\ref{pert.ratio}) because we keep $m_b$ renormalized
at $m_t$ in the top-quark contribution to $b \to s \gamma$, while all
the kinematical factors of $m_b$ are expressed in terms of the bottom
1S-mass. The bremsstrahlung function $B(E_0)$ contains the effects of
~$b \to s \gamma g$~ and ~$b \to s \gamma q \bar{q}~~(q=u,d,s)$
transitions. It is the only $E_0$-dependent part in $P(E_0)$.  Its
influence on BR$_\gamma$ is less than 4\% when $1~{\rm GeV} < E_0 <
2~{\rm GeV}$.

\newlength{\minus}
\settowidth{\minus}{$-$}
\newcommand{\m}{\hspace{\minus}}
\newlength{\zero}
\settowidth{\zero}{$0$}
\newcommand{\z}{\hspace{\zero}}
\begin{table}
\begin{center}
\caption{ Numerical results.}
\label{tab:num}
\vspace*{2mm}
\begin{tabular}{|l|l|l|l|}
\hline
                 & ``naive'' & \hspace{9mm} LO       & \hspace{8mm} NLO     \\
\hline
Re$\,K_c \; (\mu_0=M_W) $&$  -0.639  $&$  -0.631  \pm 0.003 $&$  -0.611  \pm 0.002 $\\
Re$\,K_t \; (\mu_0=m_t) $&$ \m0.450  $&$ \m0.434  \pm 0.005 $&$ \m0.397  \pm 0.003 $\\
BR$_{E_{\gamma} > 1.6\,{\rm GeV}} \times 10^4  
                        $&$ \m3.53   $&$ \m3.56\z \pm 0.14  $&$ \m3.60\z \pm 0.05  $\\
\hline
\end{tabular}
\vspace*{-5mm}
\end{center}
\end{table}

In Table~\ref{tab:num}, the numerical results are presented at various
orders of the renormalization-group-improved perturbation theory. In
the ``naive'' aprroach, the difference of $r(m_t)$ from unity is the
only included QCD effect. At LO, all the QCD logarithms ~$\left( \al
\ln M_W^2/m_b^2 \right)^n$~ are taken into account. At NLO, we add the
non-logarithmic ${\cal O}(\al)$ corrections together with the
electroweak and non-perturbative ones.  The indicated errors
correspond to varying the low-energy scale $\mu_b$ between $m_b/2$ and
$2 m_b$. One can see that the behaviour of the QCD perturbation series
for all the considered quantities is good, and that their residual
$\mu_b$-dependence is quite weak. Such a weak $\mu_b$-dependence is
not caused by any accidental cancellations, contrary to what was
observed previously.\cite{KN99} In the present approach, there is no
indication that the unknown NNLO corrections$\,$\footnote{
Except for those related to the ratio $m_c/m_b$ that has been discussed above.}
could be much larger than $(\al(m_b)/\pi)^2 \approx 0.5\%$ times a
factor of order unity. Consequently, our estimate of the overall
uncertainty in the final prediction for BR$_\gamma$ is not larger than
in the previous analyses, despite taking the problems with $m_c/m_b$
into account here.

When all the errors are included and added in quadrature, we find
\be 
{\rm BR}[\bar{B} \to X_s \gamma]^{{\rm subtracted~} \psi,\;\psi'
}_{E_{\gamma} > 1.6~{\rm GeV}}
~=~ (3.60 \pm 0.30) \times 10^{-4}.
\label{main.num}
\ee

The experimental weighted average (\ref{main.exp}) for the ``total''
branching ratio should be compared with the theoretical result
for $E_0 \simeq \f{1}{20} m_b \approx 0.23$~GeV (i.e. $\delta \equiv 1 - 2
E_0/m_b \simeq 0.9$).\cite{KN99} Then, Eq.~(\ref{main}) gives
BR$[\bar{B} \to X_s \gamma]_{E_{\gamma} > m_b/20} ~=~ 3.73 \times 10^{-4}$
with an error roughly comparable to the one in
Eq.~(\ref{main.num}). Thus, the difference between theory and
experiment is at the level of 1.4$\sigma$. However, one should
remember that the theoretical errors have no statistical
interpretation, which implies that the value of 1.4$\sigma$ has only
an illustrative character.

\end{document}